\documentclass[sigconf]{acmart}

\usepackage[utf8]{inputenc} 
\usepackage[T1]{fontenc}    
\usepackage{hyperref}       
\usepackage{url}            
\usepackage{amsfonts}       
\usepackage{microtype}      
\usepackage{float}
\usepackage{multirow}
\usepackage{graphicx}
\usepackage{xcolor}
\usepackage{colortbl}

\usepackage{amsmath}
\usepackage{amsthm}
\usepackage{enumitem}

\usepackage{xspace}

\newcommand{\framework}{XMP\xspace}
\newcommand{\method}{PaxNet\xspace}

\newtheorem{theorem}{Theorem}
\newtheorem{definition}{Definition}

\newtheorem{corollary}{Corollary}

\DeclareMathOperator*{\concat}{\scalebox{1}[1.0]{$\parallel$}}
\newcommand{\angstrom}{\text{\normalfont\AA}}

\AtBeginDocument{%
  \providecommand\BibTeX{{%
    \normalfont B\kern-0.5em{\scshape i\kern-0.25em b}\kern-0.8em\TeX}}}

\begin{document}

\title[\ ]{Efficient and Accurate Physics-aware Multiplex Graph Neural Networks for 3D Small Molecules and Macromolecule Complexes}


\author{Shuo Zhang$^{1,2}$, Yang Liu$^{2}$, Lei Xie$^{1,2,3}$}
\affiliation{%
  \institution{$^1$Ph.D. Program in Computer Science, The Graduate Center, The City University of New York \\
  $^2$Department of Computer Science, Hunter College, The City University of New York \\
  $^3$Helen \& Robert Appel Alzheimer’s Disease Research Institute, Feil Family Brain \& Mind Research Institute,  \\Weill Cornell Medicine, Cornell University \\}}
\email{szhang4@gradcenter.cuny.edu, thomasliuy@gmail.com, lxie@iscb.org}

\begin{abstract}
  Recent advances in applying Graph Neural Networks (GNNs) to molecular science have showcased the power of learning three-dimensional (3D) structure representations with GNNs. However, most existing GNNs suffer from the limitations of insufficient modeling of diverse interactions, computational expensive operations, and ignorance of vectorial values. Here, we tackle these limitations by proposing a novel GNN model, Physics-aware Multiplex Graph Neural Network (\method), to efficiently and accurately learn the representations of 3D molecules for both small organic compounds and macromolecule complexes. \method separates the modeling of local and non-local interactions inspired by molecular mechanics, and reduces the expensive angle-related computations. Besides scalar properties, \method can also predict vectorial properties by learning an associated vector for each atom. To evaluate the performance of \method, we compare it with state-of-the-art baselines in two tasks. On small molecule dataset for predicting quantum chemical properties, \method reduces the prediction error by 15$\%$ and uses 73$\%$ less memory than the best baseline. On macromolecule dataset for predicting protein-ligand binding affinities, \method outperforms the best baseline while reducing the memory consumption by 33$\%$ and the inference time by 85$\%$. Thus, \method provides a universal, robust and accurate method for large-scale machine learning of molecules. Our code is available at \url{https://github.com/zetayue/Physics-aware-Multiplex-GNN}.
\end{abstract}

\maketitle
\pagestyle{plain}

\begin{figure}[t]
  \centering
  \includegraphics[width=\columnwidth]{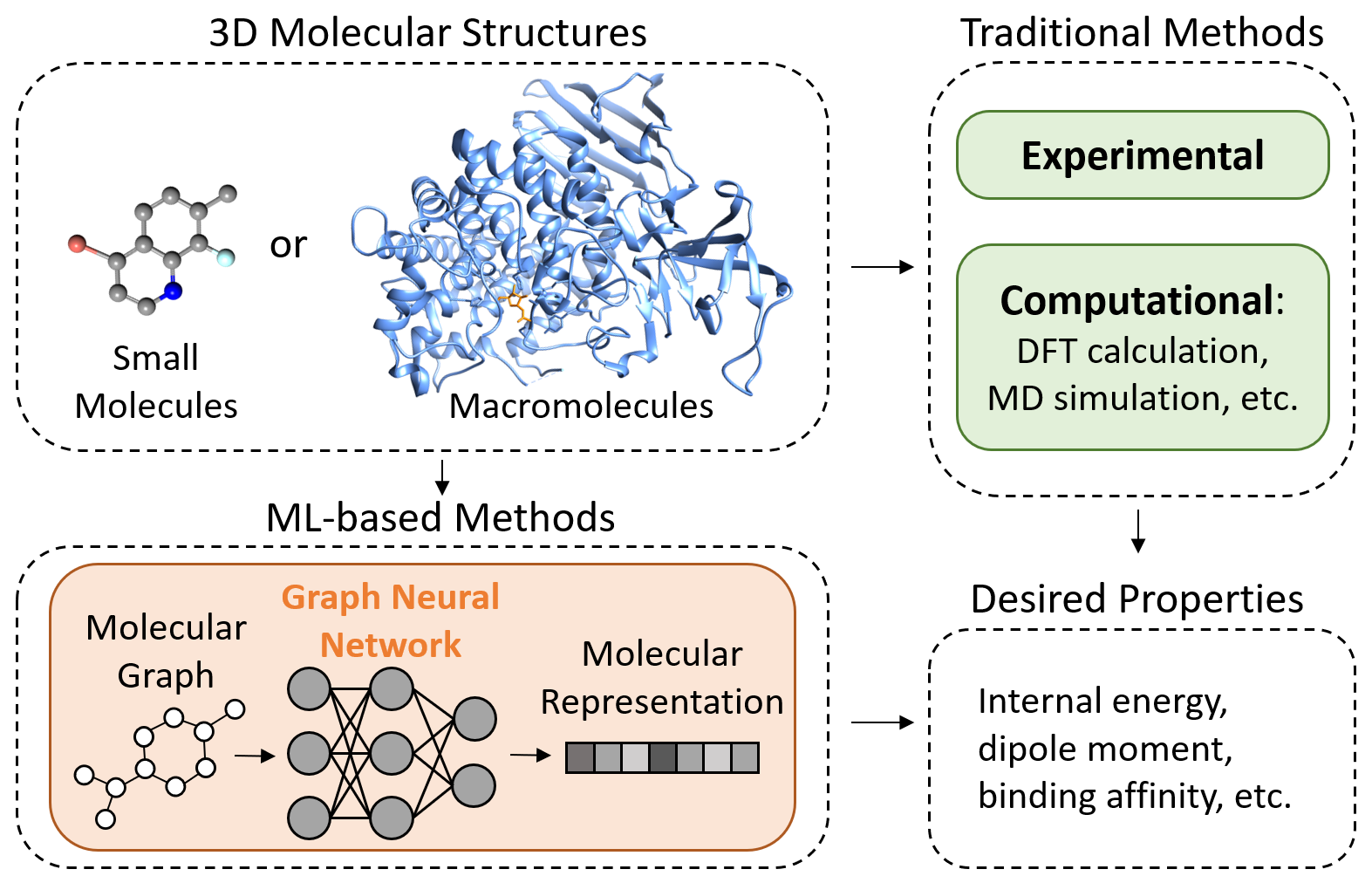}
  \vskip -0.1in
  \caption{Methods for predicting properties based on 3D molecular structures. Top left: Examples of input 3D molecular structures. Top right: Traditional methods including experimental measurements and computational approaches. Bottom left: ML-based methods especially GNNs. Bottom right: Desired properties depending on tasks.}
  \label{fig:task}
\end{figure}

\section{Introduction}
Human society benefits greatly from the discovery and design of new drug candidates with desired properties. Promising drug candidates (also known as ligands) can be small molecules that interact with macromolecules like proteins to form complexes. Such interactions modulate the cellular functions in biological processes for the treatment of diseases. As shown in Figure~\ref{fig:task}, in many related tasks like molecular property prediction and protein-ligand binding affinity prediction, 3D structures of small molecules or macromolecule complexes are used to predict the related properties including internal energy, binding affinity, etc. These properties can be traditionally either measured by experiments or computed based on simulation methods like density functional theory (DFT)~\cite{becke2014perspective} and molecular dynamics (MD)~\cite{karplus2002molecular}. However, such approaches are time-consuming. Recently, novel machine learning approaches, especially deep learning (DL) algorithms, play an increasingly important role in accelerating drug discovery~\cite{chen2018rise}. Among various DL methods, Graph Neural Networks (GNNs) have shown superior performance by treating each 3D structure as a graph and performing message passing scheme on it~\cite{wu2020comprehensive,sun2020graph,atz2021geometric,xia2021geometric,li2021structure}. Molecular graphs can naturally represent the structural information in molecules by treating atoms (or set of atoms like functional groups) as nodes, and chemical bonds (or any predefined pairwise interactions) between atoms as edges. In addition, the success of GNNs relies on the implementation of the underlying physics laws, which are symmetry and rotate invariance, associated with chemical molecules~\cite{noe2020machine,klicpera_dimenet_2020}. To better capture molecular structures and increase the expressive power of models, previous GNNs have adopted auxiliary information such as chemical properties~\cite{duvenaud2015convolutional,kearnes2016molecular,yang2019analyzing}, atomic pairwise distances in 3D space~\cite{gilmer2017neural,schutt2018schnetpack,unke2019physnet}, and angular information~\cite{klicpera_dimenet_2020,klicpera_dimenetpp_2020,shui2020heterogeneous,zhang2020molecular,li2021structure}. 

In spite of their successes, the application of GNNs in 3D representation learning for drug discovery is still in its infancy. Existing GNN models suffer from several limitations: \textbf{(1) All interactions within or between molecules are modeled by the same message passing scheme in each GNN.} Such design fails to consider the diversity of interactions (e.g. covalent or non-covalent, local or non-local) that are modeled in physics-aware approaches like molecular mechanics~\cite{schlick2010molecular}. \textbf{(2) The adoption of additional auxiliary information in GNNs requires higher computational complexity.} For example, by incorporating angular information, the resulting GNNs~\cite{klicpera_dimenet_2020,klicpera_dimenetpp_2020,shui2020heterogeneous,li2021structure} require at least $O(Nk^2)$ messages to be computed, where $N$ is the number of nodes and $k$ is the average number of nearest neighbors per node. As a comparison, the less powerful GNNs without angular information require only $O(Nk)$ messages~\cite{schutt2018schnetpack,unke2019physnet}. With restricted computational resources, those computational expensive GNNs will exhibit limited expressive power or even fail when applied to macromolecules like proteins and nucleic acids. 
\textbf{(3) Most works only focus on predicting scalar properties while ignore vectorial properties.} Although scalar properties like energy and binding affinity are essential in drug discovery, there are still many important vectorial properties like dipole moment and force. The flexibility of predicting such vectorial quantities is highly demanded in various tasks~\cite{mailoa2019fast,kouza2018role}. 

To tackle the aforementioned limitations, we propose a novel GNN framework, known as the Multiple\underline{X} \underline{M}essage \underline{P}assing (\framework), that enables the flexibility of using different message passing schemes to handle different kinds of interactions in either 2D or 3D molecules. In particular, we use multiplex graphs to represent molecular structures. Each plex (or layer) in a multiplex graph contains one type of interaction. With such input, we assign a different message passing scheme to be operated on each plex accordingly to model the interactions and update node embeddings. 

Based on our proposed \framework, we are inspired by the ideas from physics and build an efficient and accurate GNN, \underline{P}hysics-\underline{a}ware Multiple\underline{x} Graph Neural \underline{Net}work (\method), for the 3D representation learning of both small molecules and macromolecule complexes. \method takes advantage of \framework and is guided by molecular mechanics~\cite{schlick2010molecular} to model local and non-local interactions in 3D molecular structures differently. With this flexibility, \method achieves efficiency by incorporating the angular information \textit{only} in the local interactions to avoid using expensive angle-related computations on all interactions. In addition to scalar properties, \method can predict vectorial properties by capturing the geometric vectors in molecular structures that originate from quantum mechanics~\cite{veit2020predicting}.

To comprehensively verify the effectiveness of \method, we conduct experiments on benchmark datasets that involve small molecules and macromolecule complexes. For small molecules, we use QM9~\cite{ramakrishnan2014quantum} whose task is to predict quantum chemical properties of small organic molecules. For macromolecule complexes, we choose PDBbind~\cite{wang2004pdbbind} whose task is to predict the binding affinities between proteins and their small-molecule ligands. On both datasets, our model outperforms the state-of-the-art baselines in terms of not only accuracy but also efficiency regarding inference time and memory consumption. Thus, \method is suitable to large-scale machine learning for drug discovery. We summarize the main contributions as follows:
\begin{itemize}[leftmargin=10pt]
\item We propose a novel GNN framework, Multiplex Message Passing (\framework), that builds multiplex graphs to represent molecular structures and uses diverse message passing schemes to model different kinds of molecular interactions.
\item We build Physics-aware Multiplex Graph Neural Network (\method) based on \framework for the representation learning of 3D molecular structures. \method differently models the local and non-local molecular interactions with an effective and efficient design. \method is also extended to directly predict vectorial values besides scalar values.
\item Comprehensive experiments related to small molecules and macromolecule complexes are conducted to demonstrate the accuracy and efficiency of our proposed \method.
\end{itemize}

\section{Related Work}
\textbf{Graph Neural Networks for Molecular Structures.}
Graph Neural Networks (GNNs) have been proposed~\cite{duvenaud2015convolutional,niepert2016learning,kipf2017semi} to learn the representation of graph-structured data using neural networks. The architectures of GNNs implicitly encode translational, rotational, and permutation invariance, which are physics laws obeyed by molecules~\cite{noe2020machine,klicpera_dimenet_2020}. These motivate researchers to use GNNs for learning robust and generalizable representations of molecular structures in many challenging tasks such as molecular property prediction~\cite{kearnes2016molecular,gilmer2017neural,schutt2018schnetpack,klicpera_dimenet_2020}, protein-ligand binding affinity prediction~\cite{li2021structure}. Initial related works treat chemical bonds in molecules as edges and atoms as nodes to create graphs for molecules~\cite{duvenaud2015convolutional,kearnes2016molecular}. These GNNs also integrate many hand-picked chemical features to improve performance. However, they do not take account of the 3D structures of molecules, which are critical for molecular representations~\cite{townshend2020atom3d}. Thus later works turn to take atomic positions into consideration to compute interatomic distances as edge features between atoms~\cite{gilmer2017neural,schutt2017quantum,schutt2018schnetpack,unke2019physnet,liu2021transferable}. Usually, a cutoff distance is used to create molecular graphs instead of using a complete graph to reduce computational complexity and overfitting. However, GNNs may fail to distinguish certain molecules due to the choice of cutoff distance~\cite{klicpera_dimenet_2020}. To solve this issue, angular information in 3D molecular structures is further incorporated in GNNs to achieve higher expressive power~\cite{klicpera_dimenet_2020,klicpera_dimenetpp_2020, shui2020heterogeneous,li2021structure}. However, those angle-aware GNNs have significantly higher computational complexity than the previous works. With limited computational resources, they are hard to be scaled to macromolecules or large-batch learning.

Our proposed model reduces the high complexity caused by angle-related computations: As inspired by molecular mechanics, we exclude the use of angular information in the modeling of non-local interactions in molecules. Other 3D geometric-related information is carefully incorporated.

\textbf{Multiplex Graph.}
A multiplex graph (also known as a multi-view graph) consists of multiple types of edges among a set of nodes. Informally, it can be considered as a collection of graphs, where each type of edge with the same set of nodes forms a graph or a layer. To get the representation of each node, both intra-layer relationships and cross-layer relationships have to be addressed properly. In practice, various methods have been proposed to learn the embedding of the multiplex graph~\cite{zhang2018scalable,schlichtkrull2018modeling,cen2019representation} and the multiplex graph can be applied in many fields~\cite{lee2020heterogeneous,wang2020abstract}. For the representation learning on molecules, previous work~\cite{shi2020graph} implicitly represents molecular graphs as multiplex graphs and passes messages according to the edge types. Different from the existing work, we explicitly represent molecules as multiplex graphs based on the 3D geometric information in molecules. Moreover, we propose different message passing schemes for different layers in the multiplex graph.

\textbf{Neural Networks with Geometric Vectors.}
A geometric vector is a feature in $\mathbb{R}^3$ that has a direction as well as a magnitude. In the real world, geometric vectors are used to define many properties such as force, dipole moment, position vector. For the tasks related to geometric vectors, the adoption or prediction of geometric vectors is crucial in the design of neural network architectures. Especially for molecules, neural networks have been proposed to deal with geometric vectors: ~\cite{mailoa2019fast} predicts force vectors by treating them as sets of scalars and makes scalar predictions separately with expensive data augmentation to approximate rotational invariance. In other physics-informed approaches~\cite{chmiela2018towards,schutt2018schnetpack,unke2019physnet,klicpera_dimenet_2020}, the force vectors are computed by taking gradients of a predicted scalar energy field since the force field satisfies the conservation of energy. However, many geometric vectors do not have the underlying conservative scalar field. Another approach for predicting geometric vectors is to use equivariant networks with components that are all carefully designed to implement the equivariance of geometric vectors~\cite{thomas2018tensor,fuchs2020se,schutt2021equivariant}. However, for molecule-related tasks, usually the molecular representations only need to be invariant rather than equivariant. Thus the complicated equivariance may not be necessary. Recently, geometric vectors like position vectors are used to better encode the geometric features in 3D protein backbone structures~\cite{jing2020learning} using GNN. The node embeddings are updated by geometric vector features to capture more information. 

Different from the previous works, when predicting geometric vectors, we deal with general 3D molecular structures and directly make an extension based on GNN that encode invariance without complicatedly designing all components to be equivariant. Particularly, we initialize vectorial node-level contributions with position vectors to be updated by node embeddings. The finally updated vectorial node-level contributions are vector summed as the vectorial prediction.

\section{Preliminaries}
\textbf{Notations.}
Let $G = ( V , E )$ be a graph with $N=|V|$ nodes and $M = |E|$ edges. The nearest neighbors of node $i$ are defined as $\mathcal { N } ( i ) = \{ j | d ( i , j ) = 1 \}$, where $d ( i , j )$ is the shortest distance between node $i$ and $j$. The average number of the nearest neighbors of each node is $k = 2M / N$. In later formulations, we will use $\boldsymbol{h}_{i}$ as the embedding of node $i$, $\boldsymbol{e}_{j i}$ as the edge embedding between node $i$ and $j$, $\boldsymbol{m}_{ji}$ as the message being sent from node $j$ to node $i$ in the message passing scheme~\cite{gilmer2017neural}, $F$ as the hidden dimension in our model, $\mathrm{MLP}$ as multi-layer perceptron, $\concat$ as concatenation operation, $\odot$ as element wise production and $\boldsymbol{W}$ as weight matrix. 

Here we give the definition of a multiplex molecular graph, which is the input of our model, as follows:
\begin{definition} \label{def:Multiplex} \textbf{Multiplex Molecular Graph.} 
We denote a molecular structure as an $(L+1)$-tuple $G = (V, E^1, \ldots, E^L)$ where $V$ is the set of nodes (atoms) and for each $l\in\{1, 2, \ldots, L\},$ $E^l$ is the set of edges (molecular interactions) in type $l$ that between pairs of nodes (atoms) in $V$. By defining the graph $G^l = (V,E^l)$ which is also called a plex or a layer, the multiplex molecular graph can be seen as the set of graphs $G = \{G^1, G^2, ..., G^L\}$.
\end{definition}
Next we introduce the message passing scheme~\cite{gilmer2017neural} which is the basis of our model and a framework widely used in spatial-based GNNs~\cite{wu2020comprehensive}:
\begin{definition} \label{def:MP} \textbf{Message Passing Scheme.} 
Given a graph $G$, the node feature of each node $i$ is $\boldsymbol{x}_i$, and the edge feature for each node pair $j$ and $i$ is $\boldsymbol{e}_{ji}$. The message passing scheme iteratively updates message $\boldsymbol{m}_{ji}$ and node embedding $\boldsymbol{h}_{i}$ for each node $i$ using the following functions:
\begin{align}
\boldsymbol{m}_{ji}^{t} &= f_{\text {m}}(\boldsymbol{h}_{i}^{t-1}, \boldsymbol{h}_{j}^{t-1}, \boldsymbol{e}_{ji}), \\
\boldsymbol{h}_{i}^{t} &= f_{\text {u}}(\boldsymbol{h}_{i}^{t-1}, \sum\nolimits_{j \in \mathcal{N}(i)} \boldsymbol{m}_{ji}^{t}),
\end{align}
where superscript $t$ denotes the $t$-step iteration, $f_{\text {m}}$ and $f_{\text {u}}$ are learnable functions. For each node $i$, $\boldsymbol{x}_i$ is the input node embedding $\boldsymbol{h}_{i}^{0}$.
\end{definition}
In recent works~\cite{klicpera_dimenet_2020,klicpera_dimenetpp_2020,shui2020heterogeneous,li2021structure}, the message passing scheme has been modified to capture the angular information in a molecular graph $G = (V, E)$ with $N$ nodes and position vectors $\boldsymbol{r}=\left\{\boldsymbol{r}_{1}, \ldots, \boldsymbol{r}_{N}\right\}$, where $\boldsymbol{r}_{i} \in \mathbb{R}^3$ is the position of node $i$. We here analyze their computational complexity by addressing the number of angles in $G$:
\begin{theorem} \label{theo}
Given a molecular graph $G$ with position vectors $\boldsymbol{r}$, an angle is defined by a pair of adjacent edges that share a common node in $G$. By definition, there are at least $O(Nk^2)$ angles in $G$, where $N$ is the number of nodes and $k$ is the average number of nearest neighbors for each node.
\end{theorem}
Proof of Theorem~\ref{theo} can be found in Appendix~\ref{A1}. Based on Theorem~\ref{theo}, we have:
\begin{corollary} \label{corollary}
For a message passing-based GNN that requires at least one message to encode an angle, its computational complexity is at least $O(Nk^2)$ for each graph in a message passing iteration.
\end{corollary}
With Corollary~\ref{corollary}, we find that the related works~\cite{klicpera_dimenet_2020,klicpera_dimenetpp_2020,shui2020heterogeneous,li2021structure} all have at least $O(Nk^2)$ complexity.

\begin{figure*}[ht]
    \begin{center}
	\centerline{\includegraphics[width=1.6\columnwidth]{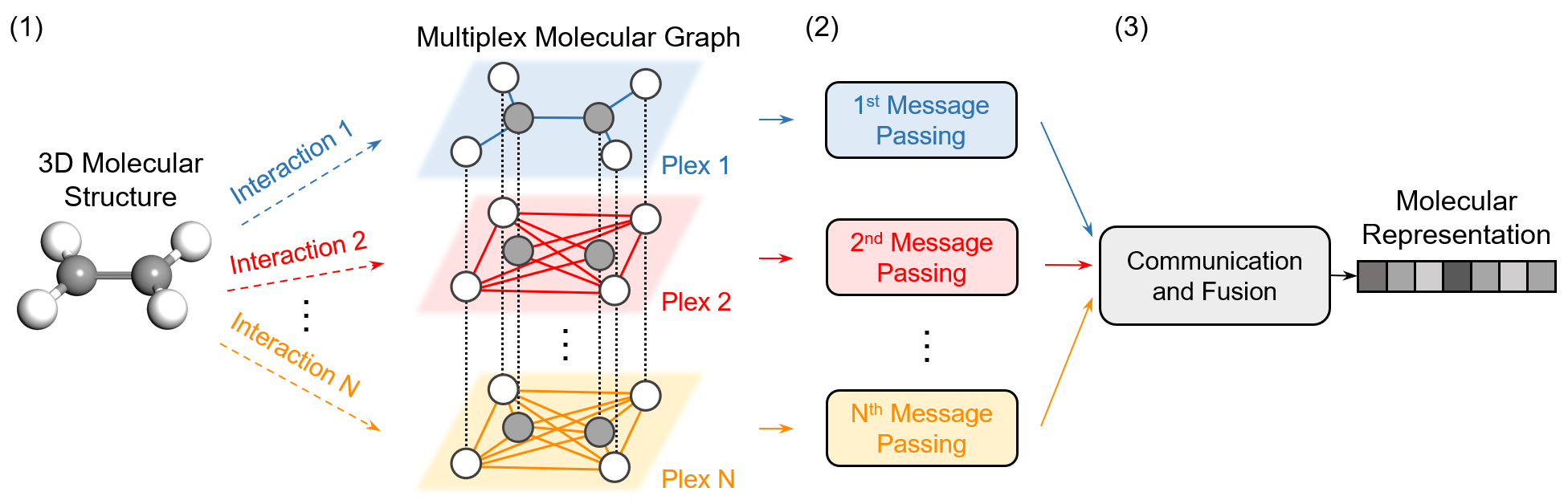}}
	\vskip -0.05in
	\caption{\label{fig:framework} Illustration of \framework framework. \framework consists of three steps including (1) the construction of multiplex molecular graph, (2) the assignment of message passing schemes, and (3) the communication and fusion of updated node embeddings.}
	\end{center}
	\vskip -0.25in
\end{figure*}

\section{Methodology}
In this section, we present \method for efficient and accurate representation learning of 3D molecular structures. We first describe the underlying \framework framework of \method. Then we describe the components of \method comprehensively.

\subsection{Multiplex Message Passing Framework}
In this work, we propose a novel GNN framework, Multiplex Message Passing (\framework), that enables the modeling of different molecular interactions with different message passing schemes to learn the representations of 2D/3D molecular structures. 

As shown in Figure~\ref{fig:framework}, \framework consists of three parts: (1) Given molecular structures, the corresponding multiplex molecular graphs are constructed based on predefined interactions in 2D/3D molecules. In detail, for each kind of interaction, we use a plex to contain them accordingly. The resulting multiplex molecular graphs are considered as input of \framework. (2) For each plex in a multiplex molecular graph, we assign a corresponding message passing scheme to model the interactions and update the node embeddings in it. (3) To address the connections across plexes, \framework communicates the updated node embeddings in different plexes. All node embeddings are finally fused together to get graph-level representations to be fed into downstream tasks.

\subsection{Physics-aware Multiplex Graph Neural Network}\label{PaxNet}
Based on \framework, we focus on learning the representations of 3D molecular structures and build Physics-aware Multiplex Graph Neural Network (\method) as an efficient and accurate approach. The design of \method is inspired by physics-based knowledge to be a novel instance of \framework.

In this section, we will first introduce the input graphs of \method, which are two-plex multiplex molecular graphs based on 3D molecular structures. Next we will present the components in \method specifically designed for the two-layer multiplex molecular graphs to learn 3D molecular representations. For vectorial property prediction, we will describe our geometric vector-based approach in detail, which is used for the prediction of dipole moments based on quantum mechanics as a practical application. Finally, we give a theoretical analysis of the computational complexity of \method to demonstrate the efficiency.

\textbf{Multiplex Molecular Graphs Inspired by Molecular Mechanics. } \label{sec:multiplex}
In molecular mechanics~\cite{schlick2010molecular}, the molecular energy $E$ is modeled with a separate consideration of local and non-local interactions: $E=E_{\text {local}}+E_{\text {nonlocal}}$. Local interactions $E_{\text {local}}=E_{\text {bond}}+E_{\text {angle}}+E_{\text {dihedral}}$ models local, covalent interactions including $E_{\text {bond}}$ that depends on bond lengths, $E_{\text {angle}}$ on bond angles, and $E_{\text {dihedral}}$ on dihedral angles. Non-local interactions $E_{\text {nonlocal}}=E_{\text {electro}}+E_{\text {vdW}}$ models non-local, non-covalent interactions including electrostatic and van der Waals interactions which depend on interatomic distances. For the geometric information in molecular mechanics, the local interactions need pairwise distances and angles, while the non-local interactions only need pairwise distances. These inspire us to avoid using expensive angle-related computations when modeling non-local interactions to achieve efficiency. 

\begin{figure*}[t]
    \centering
	\includegraphics[width=1.7\columnwidth]{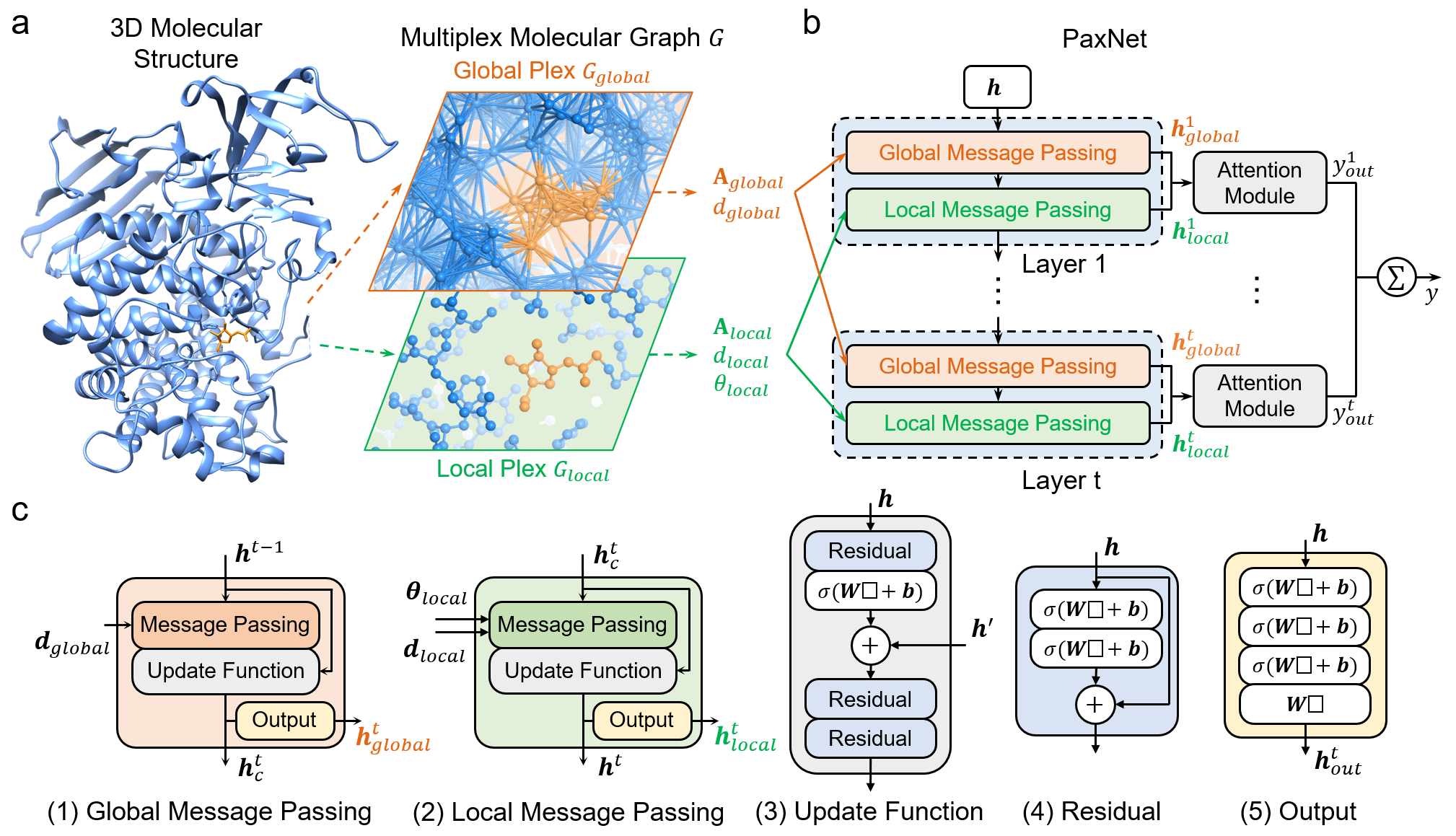}
	\vskip -0.05in
	\caption{\label{fig:method}Illustration of \method method. (a) Given a 3D molecular structure, a two-plex molecular graph $G$ is constructed to be the input of \method. (b) The overall architecture of \method. (c) The detailed components in \method.}
	\vskip -0.1in
\end{figure*}

\begin{figure*}[t]
    \begin{center}
	\centerline{\includegraphics[width=1.6\columnwidth]{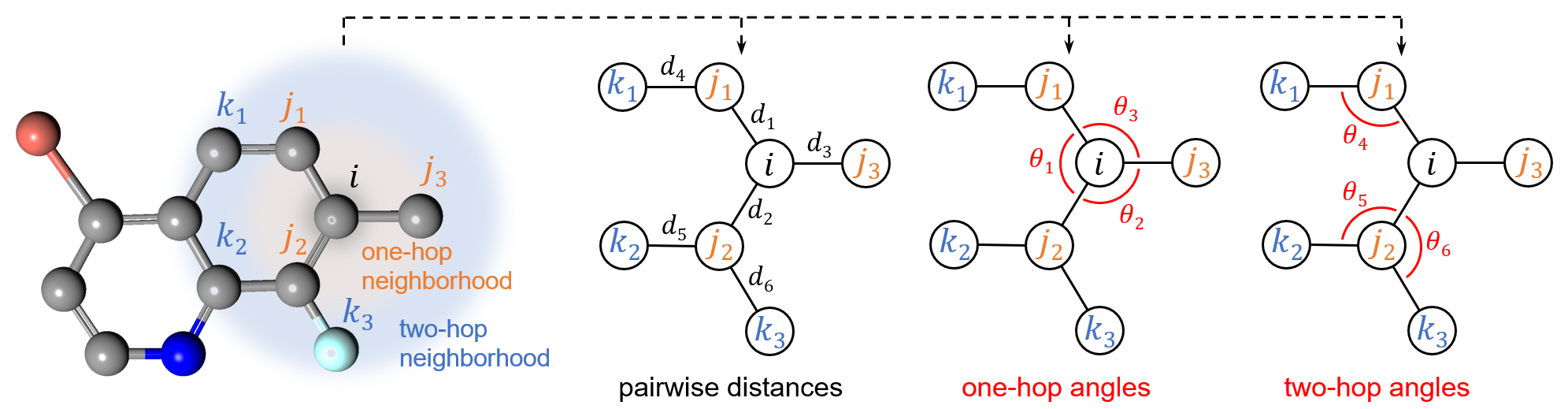}}
	\vskip -0.05in
	\caption{\label{fig:geometric} An example of the geometric information in $G$. By defining the one-hop neighbors $\{j\}$ and two-hop neighbors $\{k\}$ of node $i$, we can define the pairwise distances $d$ and the related angles $\theta$ including one-hop angles and two-hop angles.}
	\end{center}
	\vskip -0.25in
\end{figure*}

To reach our goal, we first decouple the local and non-local interactions in 3D molecular structures by using different cutoff distances when defining those interactions or just treating chemical bonds as local interactions. With the interactions, we then construct a two-plex multiplex molecular graph $G = \{G_{global}, G_{local}\}$ as shown in Figure~\ref{fig:method}a. The local plex $G_{local}$ contains only local, covalent interactions, while the global layer $G_{global}$ further includes non-local, non-covalent interactions besides local interactions. As for the geometric information, $G_{local}$ captures the related adjacency matrix $\textbf{A}_{local}$, pairwise distances $d _{local}$ and angles $\theta _{local}$, while $G_{global}$ contains only the related adjacency matrix $\textbf{A}_{global}$ and pairwise distances $d _{global}$. As illustrated in Figure~\ref{fig:geometric}, given a node $i$ in $G$, we use the geometric information within the two-hop neighborhoods around $i$.

\textbf{Overall Architecture of \method. }
As shown in Figure~\ref{fig:method}b, with the aforementioned multiplex molecular graph $G$ as input, \method uses \textit{Global Message Passing} and \textit{Local Message Passing} for $G_{global}$ and $G_{local}$ in $G$ accordingly. The node embeddings $\boldsymbol{h}$ are updated in each plex and communicated across the two plexes in $G$. To fuse $\boldsymbol{h}$ for a final representation or prediction of each molecular structure, we use an \textit{Attention Module} for each hidden layer of \method and finally sum the results together.

\textbf{Global Message Passing. }
In this module, we update node embeddings in the global plex $G_{global}$ by capturing the pairwise distances $d_{global}$ based on the message passing in Definition~\ref{def:MP}. Each message passing operation is:
\begin{align}
\boldsymbol{m}_{ji}^{t-1} &= \mathrm{MLP}_m([\boldsymbol{h}_{j}^{t-1} \concat \boldsymbol{h}_{i}^{t-1} \concat \boldsymbol{e}_{j i}]),\label{message_embedding}\\
\boldsymbol{h}_{i}^{t} &= \boldsymbol{h}_{i}^{t-1} + \sum\nolimits_{j \in \mathcal{N}(i)} \boldsymbol{m}_{ji}^{t-1}\odot \phi_{d}(\boldsymbol{e}_{j i}), \label{node_update_g}
\end{align} 
where $i, j \in G_{global}$ are connected nodes that define a message embedding, $\phi_{d}$ is a learnable function for pairwise distance. The edge embedding $\boldsymbol{e}_{j i}$ encodes the corresponding pairwise distance information. In total, this step needs $O(Nk)$ messages.

After the message passing, an update function containing multiple residual modules is used to get the node embeddings for the next layer as well as an output module for this layer. The output module is a three-layer MLP to get the output $\boldsymbol {h}_{global}$ to be fused together. Illustrations of these operations are shown in Figure~\ref{fig:method}c.

\textbf{Local Message Passing. } 
For the updates of node embeddings in the local plex $G_{local}$, we incorporate both pairwise distances $d_{local}$ and angles $\theta_{local}$. When updating the embedding of node $i$, we consider the one-hop neighbors $\{j\}$ and the two-hop neighbors $\{k\}$ of $i$. Specifically for the angles related to those nodes, we show an example in Figure~\ref{fig:geometric} and cluster them depending on the edges around $i$: (a) The \textit {one-hop angles} are angles between the one-hop edges ($\theta_{1}$, $\theta_{2}$ and $\theta_{3}$ in Figure~\ref{fig:geometric}). (b) The \textit {two-hop angles} are angles between the one-hop edges and two-hop edges ($\theta_{4}$, $\theta_{5}$ and $\theta_{6}$ in Figure~\ref{fig:geometric}). In previous GNNs that incorporate angular information, they either only address the two-hop angles~\cite{klicpera_dimenet_2020,klicpera_dimenetpp_2020} or only captures the one-hop angles~\cite{shui2020heterogeneous}. In our model, we contain all those angles to encode more related geometric information. 

To perform message passing, we use the same way as Equation (\ref{message_embedding}) to compute the message embeddings $\boldsymbol{m}$. The message passing operation in the $t$-th iteration is:
\begin{align}
\boldsymbol{m}_{ji}^{'t-1} = \boldsymbol{m}_{ji}^{t-1} &+ \sum_{j' \in \mathcal{N}(i)\setminus\{j\}} \boldsymbol{m}_{j'i}^{t-1} \odot \phi_{d}(\boldsymbol{e}_{j'i}) \odot \phi_{\theta}(\boldsymbol{\theta}_{j'i, j i}) \nonumber\\
&+ \sum_{k \in \mathcal{N}(j)\setminus\{i\}} \boldsymbol{m}_{kj}^{t-1} \odot \phi_{d}(\boldsymbol{e}_{kj}) \odot \phi_{\theta}(\boldsymbol{\theta}_{k j, j i}), \label{message_update} \\
\boldsymbol{h}_{i}^{t} = \boldsymbol{h}_{i}^{t-1} &+ \sum_{j \in \mathcal{N}(i)} \boldsymbol{m}_{ji}^{'t-1}\odot \phi_{d}(\boldsymbol{e}_{ji}) , \label{node_update_l}
\end{align}
where $i, j, k \in G_{local}$, $\phi_{d}$ is a learnable function for pairwise distance, $\phi_{\alpha}$ is a learnable function for angles. $\boldsymbol{e}_{j i}$ encodes the corresponding pairwise distance information. $\boldsymbol{\theta}_{k j, j i}$ encodes the angle $\theta_{k j, j i}=\angle k j i$ accordingly.

In Equation (\ref{message_update}), we use two summation terms to separately encode the one-hop and two-hop angles with the associated pairwise distances to update $\boldsymbol{m}_{j i}$. Motivated by Theorem~\ref{theo}, there are both $O(Nk^2)$ one-hop and two-hop angles. Thus this step needs $O(2Nk^2)$ messages. In Equation (\ref{node_update_l}), a similar operation as Equation (\ref{node_update_g}) is used to update node embedding $\boldsymbol{h}_{i}$, which requires $O(Nk)$ messages.

As shown in Figure~\ref{fig:method}c, we design the remaining functions in Local Message Passing similarly to those in Global Message Passing to get $\boldsymbol {h}_{local}$ to be fused and the input for the next iteration.

\textbf{Cross-layer Communication. }
To address the cross-layer relations between $G_{global}$ and $G_{local}$, we let the information in those layers communicate with each other as depicted in Figure~\ref{fig:method}b. We first perform Global Message Passing on $G_{global}$ in each iteration. Then the updated node embeddings are transferred to $G_{local}$ for Local Message Passing. Finally, the further updated node embeddings are passed back to $G_{global}$ for the next iteration. 

\textbf{Attention Module. }
As shown in Figure~\ref{fig:method}b, to fuse the node embeddings for a final graph-level representation or prediction as shown in the next, we design Attention Module with attention mechanism for each hidden layer $t$ of \method to get the corresponding node-level scalar prediction ${y}_{out}^t$ using $\boldsymbol {h}_{global}^t$ and $\boldsymbol {h}_{local}^t$ computed by the output modules in hidden layer $t$. 

We first compute the attention weight $\alpha_{m,i}$ that measures the contribution of $\boldsymbol {h}_{m,i}$, which belongs to node $i$ on plex $m$ in $G$:
\begin{align}
\alpha _ {m,i}^ {t} = \frac { \exp (\operatorname { LeakyReLU } (\boldsymbol{W}_m^t\boldsymbol {h}_{m,i}^t)) } { \sum _ { m} \exp ( \operatorname { LeakyReLU } (\boldsymbol{W}_m^t\boldsymbol {h}_{m,i}^t)) },\label{softmax}
\end{align}
where $m$ can be either global or local, $\boldsymbol{W}_m^t \in \mathbb{R}^{1\times F}$ is a learnable weight matrix different for each hidden layer $t$ and plex $m$. With $\alpha _ {m,i}^ {t}$, we then compute the node-level scalar prediction ${y}_{out,i}^t$ of node $i$ using weighted summation:
\begin{align}
{y}_{out,i}^t = \sum\nolimits _ { m } \alpha _ {m,i}^ {t} (\boldsymbol{W}_{out_{m}}^t\boldsymbol {h}_{m,i}^t),\label{weight_sum}
\end{align}
where $\boldsymbol{W}_{out_{m}}^t \in \mathbb{R}^{1\times F}$ is a learnable weight matrix different for each hidden layer $t$ and plex $m$. The node-level predictions ${y}_{out,i}^t$ are used to compute the final graph-level predicted result ${y}$:
\begin{align}
{y} = \sum\nolimits_{i=1}^{N}\sum\nolimits_{t=1}^{T} {y}_{out,i}^t.\label{sum}
\end{align}

\textbf{Geometric Vector-Based Approach for Predicting Vectorial Properties. }
To enable the prediction of vectorial properties instead of only scalar properties for 3D molecular structures, we design an extension of \method which can also be applied to any message-passing-based GNNs.

We extend the idea of learning scalar node-level contributions to be summed together to compute the final scalar property using GNNs. To predict vectorial properties, we aim to have \textit {vectorial node-level contributions} to be \textit {vector summed} together. We define each vectorial node-level contribution $\vec{y_i}$ to be the multiplication of the scalar atomic contribution $y_i$ and an associated vector $\vec{v}_i$. To learn $\vec{v}_i$, we propose an update function to be added in message passing operations:
\begin{align}
\vec{v}_{i}^t = f_{\vec{v}}(\boldsymbol{h}^{t}, \vec{r}),\label{vector}
\end{align}
where $\boldsymbol{h}^{t}=\{\boldsymbol{h}_{1}^{t}, \ldots, \boldsymbol{h}_{N}^{t}\}$, $\vec{r}=\{\vec{r}_{1}, \ldots, \vec{r}_{N}\}$, $\boldsymbol{h}_{i}^{t}$ is the learned embedding of node $i$ in $t$-th iteration, and $\vec{r}_{i} \in \mathbb{R}^3$ is the position vector of node $i$. In concept, $f_{\vec{v}}$ is a function that outputs $\vec{v}_i$ to be equivariant with respect to an arbitrary composition $R$ of rotations and reflections in $\mathbb{R}^3$. For example, given Equation (\ref{vector}), there should be
\begin{align}
R(\vec{v}_{i}^t)=f_{\vec{v}}(\boldsymbol{h}^t, R(\vec{r})). \label{equivariant}
\end{align}
Such constraint is necessary for vectorial property prediction and is physics-aware since the vectorial properties are also equivariant with respect to $R$ in the real world. 

To have a final predicted vectorial value $\vec{y}$, we modify Equation (\ref{weight_sum}) and (\ref{sum}) to be:
\begin{align}
\vec{y}_{out,i}^t &= \sum\nolimits _ { m } \alpha _ {m,i}^ {t} (\boldsymbol{W}_{out_{m}}^t\boldsymbol {h}_{m,i}^t)\vec{v}_{m,i}^t,\\
\vec{y} &= \sum\nolimits_{i=1}^{N}\sum\nolimits_{t=1}^{T} \vec{y}_{out,i}^t,
\end{align}
where $\vec{v}_{m,i}^t$ is the vector for node $i$ on plex $m$ in the $t$-th iteration. We multiply $\vec{v}_{m,i}^t$ with the learned scalar atomic contributions. The final prediction $\vec{y}$ is a vector sum of all vectorial contributions $\vec{y}_{out,i}^t$.

In practice, we apply our approach to predict dipole moments $\vec{\mu}$ using two different $f_{\vec{v}}$ as motivated by different quantum mechanics approximations~\cite{veit2020predicting}:

1) With the approximation that electronic charge densities are concentrated at each atomic position, we can compute $\vec{\mu}=\sum\nolimits_{i}\vec{r}_{c,i}q_i$, where $q_i$ is the partial charge of node $i$, and $\vec{r}_{c,i}=\vec{r}_i - (\sum\nolimits_{i}\vec{r}_i)/N$. We treat $q_i$ to be modeled by the scalar node-level contribution, and define $f_{\vec{v}}(\boldsymbol{h}, \vec{r})=\vec{r}_{c,i}$ to be the vector $\vec{v}_{i}$ associated with each atom $i$.

2) With the approximation of adding dipoles onto atomic positions in the distributed multipole analysis (DMA) approach~\cite{stone1981distributed}, we can write $\vec{\mu}=\sum\nolimits_{i}(\vec{r}_{c,i}q_i+\vec{\mu}_i)$, where $\vec{\mu}_i$ is the associated partial dipole of node $i$. We can rewrite the equation as $\vec{\mu}=\sum\nolimits_{i}f(\vec{r}_{i})q_i$, where $q_i$ can be modeled by the scalar node-level contribution. We treat $f(\vec{r}_{i})$ as $\vec{v}_{i}$ and define $f_{\vec{v}}(\boldsymbol{h}, \vec{r})=\sum\nolimits_{j \in \mathcal{N}(i)}|m_{i j}|(\vec{r}_{i} - \vec{r}_{j})$ to compute $\vec{v}_{i}$.

To be noted that both of the two $f_{\vec{v}}$ guarantee the equivariance of $\vec{v}_{i}$ as given by Equation (\ref{equivariant}). The reason is that $f_{\vec{v}}$ only contains linear combination of $\vec{r}_{i}$ to compute $\vec{v}_{i}$.

\textbf{Computational Complexity. }
We here analyze the computational complexity of \method by addressing the number of messages as an approximation: We denote the cutoff distance when creating the edges as $d_g$ and $d_l$ in $G_{global}$ and $G_{local}$. The average number of the nearest neighbors per node is $k_g$ in $G_{global}$ and is $k_l$ in $G_{local}$. As mentioned in previous sections, the message passings in \method require the computation of $O(Nk_g+2N{k_l}^2+Nk_l)$ messages, while previous approaches~\cite{klicpera_dimenet_2020,klicpera_dimenetpp_2020, shui2020heterogeneous,li2021structure} require $O(N{k_g}^2)$ messages. For 3D molecular structures, we have $k_g \propto {d_g}^3$ and $k_l \propto {d_l}^3$. With proper choices of $d_l$ and $d_g$, we have $k_l \ll k_g$ (e.g. $d_l=2\angstrom$ and $d_g=5\angstrom$). In such cases, our \method requires much fewer messages in message passings than those related GNNs.

\begin{table*}[t]
\caption{Performance comparison on QM9. We report the averaged results together with the standard deviations for \method. We mark the best results in bold and the second-best results with underline.}
\label{table:QM9}
\centering
\begin{tabular}{lcccccc}
\toprule
{Target} & {SchNet} & {PhysNet} & {MGCN} & {HMGNN} & {DimeNet++} & \textbf{\method}\\
\midrule
$\mu$ (D)          & \underline{0.021} & 0.0529 & 0.056 & 0.0272 & 0.0297 & \textbf{0.0108} (0.0001)\\
$\alpha$ ($a_0^3$)  & 0.124 & 0.0615 & \textbf{0.030} & 0.0561 & \underline{0.0435} & 0.0447 (0.0003)\\
$\epsilon_{\text{HOMO}}$ (meV)  & 47 & 32.9 & 42.1 & 24.78 & \underline{24.6} & \textbf{22.8} (0.3)\\
$\epsilon_{\text{LUMO}}$ (meV)  & 39 & 24.7 & 57.4 & 20.61 & \underline{19.5} & \textbf{19.2} (0.2)\\
$\Delta\epsilon$ (meV)          & 74 & 42.5 & 64.2 & 33.31 & \underline{32.6} & \textbf{31.0} (0.3)\\
$\left\langle R^{2}\right\rangle$ ($a_0^2$)  & 0.158 & 0.765 & \underline{0.11} & 0.416 & 0.331 & \textbf{0.093} (0.020)\\
ZPVE (meV)   & 1.616 & 1.39 & \textbf{1.12} & 1.18 & 1.21 & \underline{1.17} (0.02)\\
$U_0$ (meV)  & 12 & 8.15 & 12.9 & \underline{5.92} & 6.32 & \textbf{5.90} (0.12)\\
$U$ (meV)    & 12 & 8.34 & 14.4 & 6.85 & \underline{6.28} & \textbf{5.92} (0.14)\\
$H$ (meV)    & 12 & 8.42 & 16.2 & \underline{6.08} & 6.53 & \textbf{6.04} (0.14)\\
$G$ (meV)    & 13 & 9.40 & 14.6 & 7.61 & \underline{7.56} & \textbf{7.14} (0.12)\\
$c_v$ ($\frac{\mathrm{cal}}{\mathrm{mol} \mathrm{K}}$)  & 0.034 & 0.0280 & 0.038 & 0.0233 & \textbf{0.0230} & \underline{0.0231} (0.0002)\\
\midrule
std. MAE ($\%$ ) & 1.78 & 1.37 & 1.89 & 1.00 & \underline{0.98} & \textbf{0.83}\\
\bottomrule
\end{tabular}
\end{table*}

\section{Experiments}
In this section, we present our experiments using \method on two benchmark datasets to answer the following questions:
\begin{itemize}[leftmargin=10pt]
\item{\textbf{Q1}:} How accurate is our proposed \method compared with the state-of-the-art models?
\item{\textbf{Q2}:} How efficient is \method compared with the state-of-the-art models?
\item{\textbf{Q3}:} How does \method perform for predicting vectorial property?
\item{\textbf{Q4}:} Do all components in \method contribute to the performance?
\end{itemize}

\subsection{Experimental Setup}
\subsubsection{Datasets} To comprehensively evaluate the performance of \method, we use two datasets to address the 3D molecular structures in different scales from small organic molecules to protein-ligand complexes.

\textbf{QM9. }
QM9 is a widely used benchmark for the prediction of 12 molecular properties in
equilibrium~\cite{ramakrishnan2014quantum}. It consists of around 130k small organic molecules with up to 9 non-hydrogen atoms. Following~\cite{klicpera_dimenet_2020}, we randomly use 110000 molecules for training, 10000 for validation and 10831 for testing. Mean absolute error (MAE) and mean standardized MAE (std. MAE)~\cite{klicpera_dimenet_2020} are used for quantitative evaluation of the target properties.

\textbf{PDBbind. }
PDBbind is a database of experimentally measured binding affinities for protein-ligand complexes~\cite{wang2004pdbbind}. The goal is to predict the binding affinity of each protein-ligand complex based on its 3D structure. We use the same subsets in PDBbind v2016 dataset which contains ~4k structures and the same data splitting approach as in~\cite{li2021structure}. We preprocess each original complex to a structure that contains around 300 nonhydrogen atoms on average with only the ligand and the protein residues within 6$\angstrom$ around it. To comprehensively evaluate the performance, following~\cite{li2021structure}, we use Root Mean Square Error (RMSE), Mean Absolute Error (MAE), Pearson’s correlation coefficient (R), and the standard deviation (SD) in regression.

\subsubsection{Baselines}
On QM9, we compare \method with 5 baseline models: \textbf{SchNet}~\cite{schutt2018schnetpack}, \textbf{PhysNet}~\cite{unke2019physnet}, \textbf{MGCN}~\cite{lu2019molecular}, \textbf{HMGNN}~\cite{shui2020heterogeneous}, and \textbf{DimeNet++}~\cite{klicpera_dimenetpp_2020}. On PDBbind, \method is compared with 4 baselines including \textbf{D-MPNN}~\cite{yang2019analyzing}, \textbf{CMPNN}~\cite{song2020communicative}, \textbf{DimeNet}~\cite{klicpera_dimenet_2020}, and \textbf{SIGN}~\cite{li2021structure}. In all experiments, we use the original results reported in the related works for baselines. More descriptions of the baselines can be found in Appendix~\ref{A2}.

\subsubsection{Model Setup}
In our message passing operations, we define $\phi_{d}(\boldsymbol{e})=\boldsymbol{W}_{\boldsymbol{e}}\boldsymbol{e}$ and $\phi_{\alpha}(\boldsymbol{\alpha})=\mathrm{MLP}_{\alpha}(\boldsymbol{\alpha})$, where $\boldsymbol{W}_{\boldsymbol{e}}$ is a weight matrix, $\mathrm{MLP}_{\alpha}$ is a multi-layer perceptrons (MLP). For the MLPs used in our model, they all have 2 layers to take advantage of the approximation capability of MLP. For all activation functions, we use the self-gated Swish activation function as in~\cite{klicpera_dimenet_2020}. All results are reported with the averages and the standard deviations based on five random runs. More details of the settings are included in Appendix~\ref{A3}.

\subsection{Result}
\textbf{Performance on small molecule dataset (Q1). }
We show and compare \method with baseline methods on QM9 in Table~\ref{table:QM9}. \method achieves 9 best and 2 second best results among all 12 properties. It also achieves a new state-of-the-art result regarding std. MAE, which evaluates the overall performance. From the results, we can observe that the models that incorporate only atomic pairwise distance $d$ as geometric information like SchNet, PhysNet, and MGCN perform worse than those models that incorporate both $d$ and related angles $\theta$ like HMGNN, DimeNet++, and our \method. This shows the importance of capturing rich geometric information when representing 3D small molecules. 

Besides, although \method takes advantage of the concept in molecular mechanics, which is mainly used for predicting potential energies, \method works also very well for properties that are not energy-related like electronic structure properties ($\epsilon_{\text{HOMO}}$, $\epsilon_{\text{LUMO}}$, and $\Delta\epsilon$), electronic spatial extent $\left\langle R^{2}\right\rangle$, heat capacity $c_v$, and dipole moment $\mu$. The superior performance on those properties demonstrates the power of separating the modeling of different interactions in molecules based on \framework and the effectiveness of the message passing schemes involved in \method. 

\textbf{Performance on macromolecule dataset (Q1). }
On PDBbind, we compare the results of \method and baselines in Table~\ref{table:PDBBind}. \method achieves the best performance regarding all 4 evaluation metrics in our experiments. When compared with the second-best model, SIGN, \method performs significantly better (p-value < 0.05 as shown in Appendix~\ref{A4}). These results clearly demonstrate the accuracy of our model when learning representations of 3D macromolecules. The success of \method relies on the separate modeling of local and non-local interactions based on \framework. For protein-ligand complexes, the local interactions mainly capture the interactions inside a protein and a ligand, while the non-local interactions can capture the interactions between protein and ligand. \method is able to effectively handle the diverse interactions and achieve accurate results.

Among all models, D-MPNN and CMPNN only implicitly encode the geometric information in 3D structures. While DimeNet, SIGN, and our \method all explicitly use pairwise distances $d$ and related angles $\theta$. With explicitly encoded geometric information, DimeNet performs excellently on small molecule datasets. However, we find it loses superiority when compared to models designed for macromolecules like CMPNN and SIGN. This shows that DimeNet does not generalize well on macromolecules. As for SIGN which performs the best among all baselines on PDBbind, its algorithm is not directly suitable for small molecules. As a contrast, our proposed \method is generalizable for both small molecules and macromolecule complexes, and can both achieve state-of-the-art performance in our experiments.

\begin{table}[t]
\vspace{-0.5em}
\caption{Performance comparison on PDBbind. We report the averaged results together with the standard deviations. For the evaluation metrics, $\downarrow$ denotes the lower the better, while $\uparrow$ denotes the higher the better. We mark the best results in bold and the second-best results with underline.}
\label{table:PDBBind}
\centering
\small
\vskip -0.05in
\begin{tabular}{ccccc}
	\toprule
	Model & RMSE $\downarrow$ & MAE $\downarrow$ & SD $\downarrow$ & R $\uparrow$ \\
	\midrule
	D-MPNN & 1.493 (0.016) & 1.188 (0.009) & 1.489 (0.014) & 0.729 (0.006)\\
	CMPNN & 1.408 (0.028) & 1.117 (0.031) & 1.399 (0.025) & 0.765 (0.009)\\
	DimeNet & 1.453 (0.027) & 1.138 (0.026) & 1.434 (0.023) & 0.752 (0.010)\\
	SIGN & \underline{1.316} (0.031) & \underline{1.027} (0.025) & \underline{1.312} (0.035) & \underline{0.797} (0.012)\\
	\midrule
	\textbf{\method} & \textbf{1.263 (0.017)} & \textbf{0.987 (0.013)} & \textbf{1.261 (0.015)} & \textbf{0.815 (0.005)}\\
	\bottomrule
\end{tabular}
\vskip -0.05in
\end{table}

\begin{figure}[t]
  \centering
  \includegraphics[width=\columnwidth]{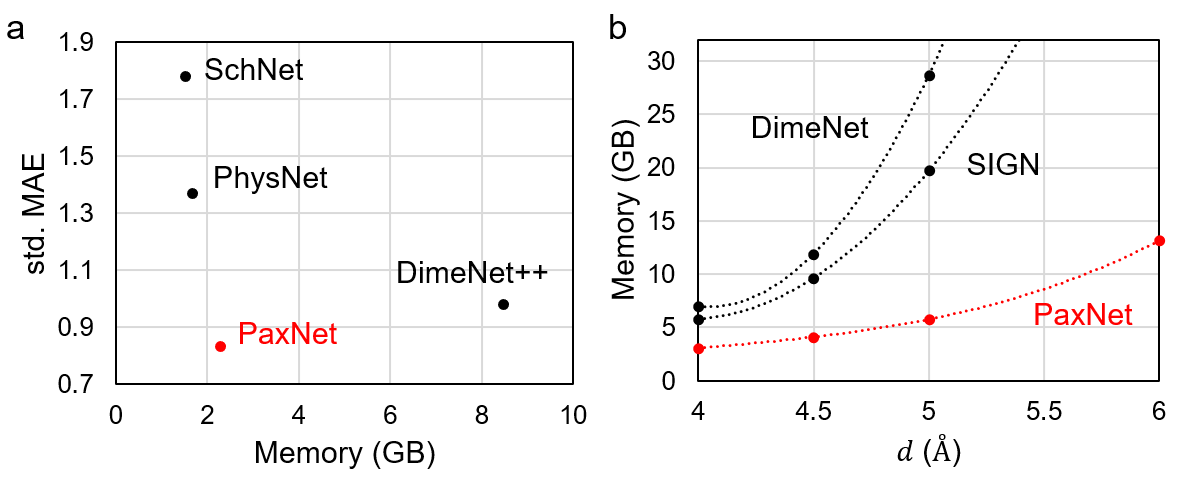}
  \vskip -0.05in
  \caption{\textbf{Results of efficiency evaluation.} (a): Std. MAE vs. memory consumption on QM9. (b): Memory consumption vs. the largest cutoff distance $d$ on PDBbind.}
  \label{fig:efficiency}
  \vskip -0.15in
\end{figure}

\textbf{Efficiency Evaluation (Q2). }
To analyze the efficiency of \method, we first use QM9 to compare \method with SchNet, PhysNet, and DimeNet++ for small molecules. The model configurations of the baselines are the same as those in their original papers. We depict the std. MAE and the memory consumption during training of each model in Figure~\ref{fig:efficiency}a. For SchNet and PhysNet, although they require less memory than \method, they perform significantly worse than \method with 114$\%$ and 65$\%$ higher std. MAE. When compared with DimeNet++, \method uses 73$\%$ less memory and reduces the std. MAE by 15$\%$. Thus \method achieves high efficiency with superior accuracy. As for the training time per epoch, \method is 26$\%$ slower than DimeNet++ by having two successive message passing schemes in each iteration. However, since \method is more memory-efficient than DimeNet++, \method can be trained with a larger batch size than DimeNet++ to achieve a speedup with the same computational resources.

We further evaluate the efficiency on PDBbind to compare \method with DimeNet and SIGN that explicitly use 3D information for macromolecules. In macromolecules, the cutoff distance $d$ plays an important role to capture inter-molecular interactions~\cite{li2021structure} and directly affects the computational complexity. Thus we compare the memory consumption when using different $d$ in related models in Figure~\ref{fig:efficiency}b. From the results, we observe that the memory consumed by DimeNet and SIGN increases much faster than \method when $d$ increases. When fixing $d=5\angstrom$ as an example, \method requires 80$\%$ and 71$\%$ less memory than DimeNet and SIGN, respectively. Thus \method is much more memory-efficient and is able to capture longer-range interactions than these baselines with restricted resources. When comparing SIGN with \method, both of which use the configurations to achieve the results in Table~\ref{table:PDBBind}, we find that \method outperforms SIGN while reducing the memory consumption by 33$\%$ and the inference time by 85$\%$. This clearly demonstrates the accuracy and efficiency of \method.

\textbf{Performance on Vectorial Property Prediction (Q3). }
On QM9 when predicting dipole moment $\mu$, we can simply predict $\mu$ as a scalar property, or predict the original vectorial value $\vec{\mu}$ and get its magnitude: $\mu = |\vec{\mu}|$. We compare three different methods of using \method to predict $\mu$: 1) No geometric vectors are used in \method. 2) Geometric vectors are defined as $\vec{v}=\vec{r}_{c,i}$. 3) Geometric vectors are defined as $\vec{v}=\sum\nolimits_{j \in \mathcal{N}(i)}|m_{i j}|(\vec{r}_{i} - \vec{r}_{j})$. Method 1 directly predict scalar values, while method 2 and 3 are those we proposed in Section~\ref{PaxNet} and can directly predict vectorial values.

Among the three methods, method 3 achieves the lowest MAE with 0.0108D, which significantly outperforms the best baseline result (0.021D) in Table~\ref{table:QM9}. Method 2 gets an MAE of 0.0120D, which is still much better than the other baselines. On the contrary, the scalar-based method 1 get 0.0240D which is the worst one among our three methods based on \method. The success of method 3 relies on our geometric vector-based approach and better physics-aware approximation compared to the one used in method 2.

\begin{table}[t]
\caption{Ablation study of \method on QM9. We compare the variants with the original \method and report the differences of MAE on different targets.} \label{table:ablation}
\vskip -0.05in
\centering
\small
\newcommand{\tabincell}[2]{\begin{tabular}{@{}#1@{}}#2\end{tabular}}
\begin{tabular}{lrr}
	\toprule
	\multirow{2}{*}{\method Variant} & \multicolumn{2}{c}{MAE on Target}\\
	 & $U_0$, $U$, $H$, $G$ & $\mu$\\
	\midrule
	No Attention Module   & +4\% & +3\%\\
	No Communication between MPs   & +10\% & +21\%\\
    Global MP + Local MP (One-hop $\theta$) & +4\% & +8\%\\
    Global MP + Local MP (Two-hop $\theta$) & +6\% & +11\%\\
	Global MP   & +28\% & +29\%\\
    Local MP (all $\theta$)  & +49\% & +102\%\\
    Local MP (One-hop $\theta$)  & +57\% & +107\%\\
    Local MP (Two-hop $\theta$)  & +64\% & +111\%\\
	\bottomrule
\end{tabular}
\vskip -0.05in
\end{table}

\textbf{Ablation Study for (Q4). }
To test whether all of the components in \method, including attention module, message passing schemes, and communications across plexes, contribute to the success of \method, we conduct ablation study on QM9 by designing \method variants: Without the attention module, we use an average of all output results. By removing the communications across plexes, we perform the message passing on two plexes in parallel without communication. We also remove either Global Message Passing or Local Message Passing. For Local Message Passing, we design variants by considering different $\theta$. The performances of all variants are evaluated on scalar targets including $U_0$, $U$, $H$ and $G$, and vectorial target $\mu$. The results in Table~\ref{table:ablation} show that all variants decrease the performance of the original \method. These results validate the contributions of those components. For $\theta$, we find that using all $\theta$ performs the best. It demonstrates the importance of addressing more geometric information.

\section*{Conclusion}
In this work, we propose \framework, which is a novel GNN framework to better address the diverse molecular interactions with different message passing schemes than state-of-the-art algorithms. Based on \framework, we are inspired by the ideas in physics and propose to separate the modeling of local and non-local interactions in molecules. Such design makes it possible to effectively model the interactions and avoid many computational expensive operations. The resulting GNN, \method, can also predict vectorial properties by learning an associated vector for each node. The experiments conducted that involve both small molecules and macromolecule complexes clearly demonstrate the efficiency and accuracy of \method.

\bibliographystyle{ACM-Reference-Format}
\bibliography{main}


\newpage
\appendix

\section{Appendix}

\subsection{Proof of Theorem~\ref{theo}} \label{A1}
\begin{proof} 
Given a graph $G$ with $N$ nodes, the $i$-th node has a degree $k_i$ for each $i \in N$. The mean degree $k=\sum\nolimits _ {i}{k_i}/N$. Since every two connected edges can define one angle, we can define $(k_i(k_i-1))/2$ angles for node $i$ with $k_i$ edges connected to it. Thus the total number of angles in $G$ is $O(\sum\nolimits _ {i}{k_i^2})$. From Cauchy–Schwarz inequality, we know that $\sum\nolimits _ {i}{k_i^2} \leq (\sum\nolimits _ {i}{k_i})^2/N = Nk^2$. So that the number of angles in $G$ is at least $O(Nk^2)$.
\end{proof}

\subsection{Baseline Descriptions} \label{A2}
\begin{itemize}[leftmargin=10pt]
\item \textbf{SchNet}~\cite{schutt2018schnetpack} is a GNN that uses continuous-filter convolutional layers to model atomistic systems. Interatomic distances are used when designing convolutions.
\item \textbf{PhysNet}~\cite{unke2019physnet} uses message passing scheme for predicting properties of chemical systems. It models chemical interactions with learnable distance-based functions.
\item \textbf{MGCN}~\cite{lu2019molecular} utilizes the multilevel structure in molecular system to learn the representations of quantum interactions level by level based on GNN. The final molecular property prediction is made with the overall interaction representation.
\item \textbf{HMGNN}~\cite{shui2020heterogeneous} is a GNN-based method for predicting molecular properties. It represents molecules with heterogeneous molecular graphs to model many-body interactions. Its message passing scheme leverages both distances and angles as features.
\item \textbf{DimeNet}~\cite{klicpera_dimenet_2020} is a message passing neural network using directional message passing scheme for small molecules. Both distances and angles are used when modeling the molecular interactions.
\item \textbf{DimeNet++}~\cite{klicpera_dimenetpp_2020} is an improved version of DimeNet with better accuracy and faster speed. It can also be used for non-equilibrium molecular structures.
\item \textbf{D-MPNN}~\cite{yang2019analyzing} is a message passing neural network that incorporates edge features. The aggregation process addresses the pairwise distance information contained in edge features.
\item \textbf{CMPNN}~\cite{song2020communicative} is built based on D-MPNN and has a communicative message passing scheme between nodes and edges for better performance when learning molecular representations.
\item \textbf{SIGN}~\cite{li2021structure} is a recent state-of-the-art GNN for predicting protein-ligand binding affinity. It builds complex interaction graphs for protein-ligand complexes and integrates both distance and angle information in modeling.
\end{itemize}

\subsection{Experimental Setting Details} \label{A3}
\subsubsection{Sources of Datasets}
In this section, we present the sources of downloading the datasets used in our experiments and more details of the datasets.

\textbf{QM9. }
For QM9 dataset, we use the source\footnote{\url{https://deepchemdata.s3-us-west-1.amazonaws.com/datasets/molnet_publish/qm9.zip}} provided by MoleculeNet~\cite{wu2018moleculenet}. Following the previous works~\cite{klicpera_dimenet_2020,klicpera_dimenetpp_2020, shui2020heterogeneous}, we process QM9 by removing about 3k molecules that fail a geometric consistency check or are difficult to converge. For properties $U_0$, $U$, $H$, and $G$, only the atomization energies are used by subtracting the atomic reference energies as in~\cite{klicpera_dimenet_2020,klicpera_dimenetpp_2020, shui2020heterogeneous}. For property $\Delta \epsilon$, we follow the same way as the DFT calculation and predict it by calculating $\epsilon_{\mathrm{LUMO}}-\epsilon_{\mathrm{HOMO}}$.

\textbf{PDBbind. }
For PDBbind dataset, we use the same data source\footnote{\url{https://github.com/PaddlePaddle/PaddleHelix/tree/dev/apps/drug_target_interaction/sign}} as used in SIGN~\cite{li2021structure}. We use log$K_i$ as the target property being predicted, which is proportional to the binding free energy.

\subsubsection{Sources of Model Implementations}
To make fair comparisons and exclude the differences brought by different frameworks when performing efficiency evaluation in our experiments, we adopt or build the implementations of the related models based on PyTorch: For SchNet\footnote{\url{https://github.com/rusty1s/pytorch_geometric/blob/73cfaf7e09/examples/qm9_schnet.py}}~\cite{schutt2018schnetpack} and DimeNet\footnote{\url{https://github.com/rusty1s/pytorch_geometric/blob/73cfaf7e09/examples/qm9_dimenet.py}}~\cite{klicpera_dimenet_2020}, we adopt the implementations provided by PyTorch Geometric library\footnote{\url{https://github.com/pyg-team/pytorch_geometric}}. For PhysNet~\cite{unke2019physnet} and DimeNet++~\cite{klicpera_dimenetpp_2020}, since PhysNet, DimeNet, and DimeNet++ use similar framework when building their models, we build PhysNet and DimeNet++ based on the aforementioned implementation of DimeNet. For SIGN~\cite{li2021structure}, we use the original PyTorch-based implementation mentioned in its work.

\begin{table}[t]
\caption{List of hyperparameters used in our experiments.}
\label{table:hyperparameter}
\centering
\vskip -0.05in
\begin{tabular}{lcc}
	\toprule
	\multirow{2}{*}{Hyperparameters} & \multicolumn{2}{c}{Value}\\
	  & QM9 & PDBbind\\
	\midrule
	Batch Size & 32, 128 & 32\\
	Hidden Dim. & 128 & 128\\
	Initial Learning Rate & 1e-4 & 1e-3\\
	Number of Layers & 6 & 3\\
	Max. Number of Epochs & 900 & 100\\
	\bottomrule
\end{tabular}
\vskip -0.05in
\end{table}

\begin{table*}[t]
\vspace{-0.5em}
\caption{Statistical significance (p-value) between \method and SIGN on PDBbind. The best results are marked in bold.}
\label{table:significance}
\centering
\vskip -0.05in
\begin{tabular}{ccccc}
	\toprule
	Model & RMSE $\downarrow$ & MAE $\downarrow$ & SD $\downarrow$ & R $\uparrow$ \\
	\midrule
	SIGN & 1.316 (0.031) & 1.027 (0.025) & 1.312 (0.035) & 0.797 (0.012)\\
	\textbf{\method} & \textbf{1.263 (0.017)} & \textbf{0.987 (0.013)} & \textbf{1.261 (0.015)} & \textbf{0.815 (0.005)}\\
	\midrule
	Significance (p-value) & 0.0122 & 0.0156 & 0.0242 & 0.0212\\
	\bottomrule
\end{tabular}
\vskip -0.05in
\end{table*}

\subsubsection{Implementation Details}
In this section, we describe more details of the implementation of \method in our experiments. All of the experiments are done on an NVIDIA Tesla V100 GPU (32 GB).

\textbf{QM9. }
In the multiplex molecular graphs for QM9, we use chemical bonds as the edges in the local plex, and a cutoff distance (5 or 10$\angstrom$) to create the edges in the global plex. The 3D molecular structures are processed using the RDKit library. For the input features, we use atomic numbers $Z$ as node features and pairwise distances between atoms as edge features computed from atomic positions $\vec{r}$. Following~\cite{schutt2018schnetpack,unke2019physnet,klicpera_dimenet_2020}, we represent $Z$ with randomly initialized, trainable embeddings and expand pairwise distances and angles with basis functions to reduce correlations. For the basis functions $\boldsymbol{e}_{RBF}$ and $\boldsymbol{a}_{SBF}$, we use $N_{\text{SHBF}}=7$, $N_{\text{SRBF}}=6$ and $N_{\text{RBF}}=16$. In our experiment, we use the single-target training following~\cite{klicpera_dimenet_2020} by using a separate model for each target instead of training a single shared model for all targets. The models are optimized by minimizing the mean absolute error (MAE) loss using the Adam optimizer. We use a linear learning rate warm-up over 1 epoch and an exponential decay with ratio 0.1 every 600 epochs. The model parameter values for validation and test are kept using an exponential moving average with a decay rate of 0.999. To prevent overfitting, we use early stopping on the validation loss. For properties ZPVE, $U_0$, $U$, $H$, and $G$, we use the cutoff distance in the global layer $d_g=5\angstrom$. For the other properties, we use $d_g=10\angstrom$.

\textbf{PDBbind. }
In our multiplex molecular graphs for PDBbind, we use cutoff distance $d_l=2\angstrom$ for the local plex and $d_g>2\angstrom$ for the global plex. For the input features, we use 18 atom features as used in~\cite{li2021structure} to be node features. The edge features and the basis functions used to expand pairwise distances and angles are the same as those used on QM9. For each model being investigated, we create three weight-sharing, replica networks, one each for predicting the target value $y$ of protein-ligand complex, protein pocket, and ligand following~\cite{gomes2017atomic}. The final target $\Delta y$ is computed by $\Delta y = y_{\text{complex}} - y_{\text{pocket}} - y_{\text{ligand}}$. The full model is trained by minimizing the mean squared error (MSE) loss between $\Delta y$ and the true values using the Adam optimizer. The learning rate is dropped by a factor of 0.2 every 50 epoches. The validation losses are used for early stopping.

In Table~\ref{table:hyperparameter}, we list the most important hyperparameters used in our experiments.

\subsection{Statistical significance between SIGN and \method on PDBbind} \label{A4}
We use p-value to compute the statistical significance between SIGN and \method on PDBbind. As shown in Table~\ref{table:significance}, \method performs significantly better than SIGN on all four metrics with p-value < 0.05.

\end{document}